\documentclass[aps,twocolumn]{revtex4}
\usepackage{graphicx}
\usepackage{epsfig}
\usepackage{dcolumn}
\usepackage{amsmath}
\def\Journal#1#2#3#4{{#1} {\bf#2}, #3 (#4)}

\def\PLB{{\rm Phys. Lett.}  B}
\def\PRL{\rm Phys. Rev. Lett.}
\def\PRD{{\rm Phys. Rev.} D}

\def\JPG{{\rm J. Phys.} G}

\def\ep{\epsilon}
\def\lam{\lambda}

\def\la{\langle}
\def\ra{\rangle}

\def\al{\alpha}

\def\be{\begin{equation}}
\def\ee{\end{equation}}
\def\bea{\begin{eqnarray}}
\def\eea{\end{eqnarray}}
\input{epsfig.sty}
\begin{document}
\title{Decay constants and radiative decays of heavy mesons 
in light-front quark model}
\author{ Ho-Meoyng Choi\\
Department of Physics, Teachers College, Kyungpook National University,
     Daegu, Korea 702-701}
\begin{abstract}
We investigate the magnetic dipole decays $V\to P\gamma$ 
of various heavy-flavored mesons such as $(D,D^*,D_s,D^{*}_s,\eta_c, J/\psi)$
and $(B,B^*,B_s,B^*_s,\eta_b,\Upsilon)$ using the light-front quark model 
constrained by the variational principle for the QCD-motivated effective
Hamiltonian. The momentum dependent form factors $F_{VP}(q^2)$ for
$V\to P\gamma^*$ decays are obtained
in the $q^+=0$ frame and then analytically continued to the timelike region
by changing ${\bf q}_\perp$ to $i{\bf q}_\perp$ in the form factors.
The coupling constant $g_{VP\gamma}$ for real photon case is then obtained
in the limit as $q^2\to 0$, i.e. $g_{VP\gamma}=F_{VP}(q^2=0)$.
The weak decay constants of heavy pseudoscalar
and vector mesons are also calculated. Our numerical results for the 
decay constants and radiative decay widths for the heavy-flavored mesons
are overall in good agreement with the available experimental data as
well as other theoretical model calculations. 
\end{abstract}


\maketitle
\section{Introduction}
The physics of exclusive heavy meson decays
has provided very useful testing ground for the precise 
determination of the fundamental parameters of the standard model(SM) and 
the development of a better understanding of the QCD dynamics.
While the experimental tests of exclusive heavy meson decays are 
much easier than those of inclusive one, the theoretical understanding 
of exclusive decays is complicated mainly due to the nonperturbative
hadronic matrix elements entered in the long distance nonperturbative
contributions. Since a rigrous field-theoretic formulation
with a first principle application of QCD to make a 
reliable estimates of the nonperturbative hadronic matrix elements
has not so far been possible, most of theoretical efforts have been 
devoted to looking for phenomenological approaches to nonperturbative 
QCD dynamics.

Along with various exclusive processes such as leptonic, semileptonic
and rare decays of heavy mesons, the one-photon radiative 
decays from the low-lying heavy vector(V) to heavy pseudoscalar(P) mesons, 
i.e. magnetic dipole $V(1^3S_1)\to P(1^1S_0)\gamma$ transitions,
have been considered as a valuable testing ground 
to pin down the best phenomenological model of hadrons. For example,
the calculations of $D^*\to D\gamma$ and $B^*\to B\gamma$ radiative decays
have been investigated by various theoretical approaches, such as
the quark model~\cite{GI,Barik,Jaus96,Jones,EFG1}, light cone QCD sum 
rules~\cite{Dosc,Aliev}, heavy quark effective theory(HQET)~\cite{Cheng,HQET},
cloudy bag model(CBM)~\cite{CBM},
and chiral perturbation theory~\cite{Amu}. Recently, the radiative 
decays between two heavy  quarkonia such as $J/\psi\to\eta_c\gamma$ and 
$\Upsilon\to\eta_b\gamma$ have also been studied by the potential 
nonrelativistic QCD(pNRQCD)~\cite{Nora} and relativistic quark
model~\cite{Hwang,EFG2} approaches. In our previous light-front quark
model(LFQM) analysis~\cite{CJ1,CJ2,CJ3} 
based on the QCD-motivated effective Hamiltonian, we have 
analyzed various exlcusive processes such as the $0^-\to0^-$ semileptonic
heavy meson decays~\cite{CJ2}, rare $B\to Kl^+l^-(l=e,\mu,\tau)$ 
decays~\cite{CJ3} and radiative $V\to P\gamma$ and 
$P\to V\gamma$ decays of light-flavored 
mesons($\pi,\rho,\omega,K,K^*,\phi,\eta,\eta'$)~\cite{CJ1}
and found a good agreement with the experimental data.

The main purpose of this work is to 
investigate the magnetic dipole transition $V\to P\gamma$
for the heavy-flavored mesons such as $(D,D^*,D_s,D^{*}_s,\eta_c, J/\psi)$ and
$(B,B^*,B_s,B^*_s,\eta_b,\Upsilon)$ using our LFQM~\cite{CJ1,CJ2,CJ3}. 
Since the experimental data available 
in this heavy-flavored sector are scanty, 
predictions of a model, if found reliable, can be utilized quite
fruitfully. In addition, we calculate the weak decay constants of heavy
pseudoscalar and vector mesons, which play important roles in many aspects,
such as in the determination of Cabibbo-Kobayashi-Maskawa matrix elements,
in the leptonic or nonleptonic weak decays of mesons, and in the neutral
$D-\bar{D}$ or $B-\bar{B}$ mixing process, etc.
Our LFQM~\cite{CJ1,CJ2,CJ3} used in the present analysis 
has a couple of salient features compared to other LFQM~\cite{Jaus96,Hwang} 
analysis: (1) We have implemented the variational principle to QCD-motivated 
effective LF Hamiltonian to enable us to analyze the meson mass spectra
and to find optimized model parameters, which are to be used subsequently 
in the present investigation. Such an approach can better constrain 
the phenomelogical parameters and  establish the extent of applicability 
of our LFQM to wider ranging hadronic phenomena.
(2) We have performed the analytic continuation from the 
spacelike($q^2<0$) region to the physical timelike 
region($0\leq q^2\leq(M_V-M_P)^2$) to obtain the decay form 
factors $F_{VP}(q^2)$ for $V\to P\gamma^*$ transitions. The 
Drell-Yan-West($q^+=q^0+q^3=0$) frame is useful because only valence 
contributions are needed, i.e. the hadronic matrix element 
$\la P|J^\mu_{\rm em}|V\ra$ is represented
as the overlap of valence wave function,
as far as the ``$\mu=+$" component of the current is used. 

The paper is organized as follows: 
In Sec.II, we briefly describe the formulation of our LFQM~\cite{CJ1,CJ2}
and the procedure of fixing the model parameters using the variational
principle for the QCD-motivated effective Hamiltonian. The decay constants
and radiative $V\to P\gamma$ decay widths for heavy-flavored mesons are
then uniquely determined in our model calculation.
In Sec. III, the formulae for the decay constants of pseudoscalar
and vector mesons as well as the decay widths for $V\to P\gamma$ in our
LFQM are given. To obtain the $q^2$-dependent transition form factors
$F_{VP}(q^2)$ for $V\to P\gamma^*$ transitions, we use the $q^+=0$ frame(i.e.
$q^2=-{\bf q}^2_\perp<0$) and then analytically continue the spacelike
results to the timelike $q^2>0$ region by changing ${\bf q}_\perp$ to 
$i{\bf q}_\perp$ in the form factor. The coupling constants
$g_{VP\gamma}$ needed for the calculations of the decay widths for
$V\to P\gamma$ can then be determined in the limit as 
$q^2\to 0$, i.e. $g_{VP\gamma}=F_{VP}(q^2=0)$.
In Sec. IV, we present our numerical results and compare
with the available experimental data as well as other theoretical 
model predictions. Summary and conclusions follow in Sec.V.
 
\section{Model Description}
The key idea in our LFQM~\cite{CJ1,CJ2} for mesons is to treat the radial 
wave function as trial function for the variational principle to the 
QCD-motivated effective Hamiltonian saturating the Fock state expansion 
by the constituent quark and antiquark. The QCD-motivated Hamiltonian for 
a description of the ground state meson mass spectra is given by
\bea\label{Ham}
H_{q\bar{q}}|\Psi^{JJ_z}_{nlm}\ra&=&\biggl[
\sqrt{m^2_q+{\vec k}^2}+\sqrt{m^2_{\bar{q}}+{\vec k}^2}+V_{q\bar{q}}\biggr]
|\Psi^{JJ_z}_{nlm}\ra,
\nonumber\\
&=&[H_0 + V_{q\bar{q}}]|\Psi^{JJ_z}_{nlm}\ra
=M_{q\bar{q}}|\Psi^{JJ_z}_{nlm}\ra,
\eea
where ${\vec k}=({\bf k}_\perp, k_z)$ is the three-momentum of the 
constituent quark, $M_{q\bar{q}}$ is the mass 
of the meson, and $|\Psi^{JJ_z}_{nlm}\ra$ is the meson
wave function. In this work, we use two interaction potentials $V_{q\bar{q}}$
for the pseudoscalar($0^{-+}$) and vector($1^{--}$) mesons: (1) Coulomb
plus harmonic oscllator(HO), and (2) Coulomb plus linear confining potentials.
In addition, the hyperfine interaction, which is essential to distinguish
vector from pseudoscalar mesons, is included for both cases, viz.,
\be\label{pot}
V_{q\bar{q}}=V_0 + V_{\rm hyp}
= a + {\cal V}_{\rm conf}-\frac{4\al_s}{3r}
+\frac{2}{3}\frac{{\bf S}_q\cdot{\bf S}_{\bar{q}}}{m_qm_{\bar{q}}}
\nabla^2V_{\rm coul},
\ee
where ${\cal V}_{\rm conf}=br(r^2)$ for the linear(HO) potential and
$\la{\bf S}_q\cdot{\bf S}_{\bar{q}}\ra=1/4(-3/4)$ for the 
vector(pseudoscalar) meson.

The momentum space light-front wave function of the ground state
pseudoscalar and vector mesons is given by 
\be\label{w.f}
\Psi^{JJ_z}_{100}(x_i,{\bf k}_{i\perp},\lam_i)
={\cal R}^{JJ_z}_{\lam_1\lam_2}(x_i,{\bf k}_{i\perp})
\phi(x_i,{\bf k}_{i\perp}),
\ee
where $\phi(x_i,{\bf k}_{i\perp})$ is the radial wave function and 
${\cal R}^{JJ_z}_{\lam_1\lam_2}$ is the spin-orbit wave function,
which is obtained by the interaction independent Melosh transformation
from the ordinary equal-time static spin-orbit wave function assigned
by the quantum numbers $J^{PC}$.
The model wave function in Eq.~(\ref{w.f}) is represented by the
Lorentz-invariant variables, $x_i=p^+_i/P^+$, 
${\bf k}_{i\perp}={\bf p}_{i\perp}-x_i{\bf P}_\perp$ and $\lam_i$, where
$P^\mu=(P^+,P^-,{\bf P}_\perp)
=(P^0+P^3,(M^2+{\bf P}^2_\perp)/P^+,{\bf P}_\perp)$ is the momentum of the
meson $M$, $p^\mu_i$ and $\lam_i$ are the momenta and the helicities of 
constituent quarks, respectively.

The covariant forms of the spin-orbit wave functions
for pseudoscalar and vector mesons are given by 
\bea\label{R00_A}
{\cal R}_{\lam_1\lam_2}^{00}
&=&\frac{-\bar{u}(p_1,\lam_1)\gamma_5 v(p_2,\lam_2)}
{\sqrt{2}\tilde{M_0}},
\nonumber\\
{\cal R}_{\lam_1\lam_2}^{1J_z}
&=&\frac{-\bar{u}(p_1,\lam_1)
\biggl[/\!\!\!\ep(J_z) -\frac{\ep\cdot(p_1-p_2)}{M_0 + m_1 + m_2}\biggr]
v(p_2,\lam_2)} {\sqrt{2}\tilde{M_0}},
\nonumber\\
\eea
where $\tilde{M_0}=\sqrt{M^2_0-(m_1-m_2)^2}$ 
and $M^2_0$  is the invariant meson mass square $M^2_0$ defined as 
\be\label{IM}
M^2_0=\sum_{i=1}^2\frac{{\bf k}^2_{i\perp}+m^2_i}{x_i}.
\ee
The polarization vectors $\ep^\mu(J_z)$ of the vector meson with four
momentum $P$ are given by 
\bea{\label{pol_vec}}
\epsilon^\mu(\pm 1)&=&
\biggl[0,\frac{2}{P^+}{\bf\epsilon}_\perp(\pm)\cdot{\bf P_{\perp}},
{\bf\epsilon}_\perp(\pm 1)\biggr],
\nonumber\\
{\bf\epsilon}_\perp(\pm 1)&=&\mp\frac{(1,\pm i)}{\sqrt{2}},
\nonumber\\
\epsilon^\mu(0)&=&
\frac{1}{M_0}\biggl[P^+,\frac{{\bf P}^2_{\perp}-M^2_0}{P^+},
{\bf P}_{\perp}\biggr].
\eea
The spin-orbit wave functions satisfy the following relations
\bea\label{Rsum}
\sum_{\lam_1\lam_2}{\cal R}_{\lam_1\lam_2}^{JJ_z\dagger}
{\cal R}_{\lam_1\lam_2}^{JJ_z}=1,
\eea
for both pseudoscalar and vector mesons.
For the radial wave function $\phi$, we use the same Gaussian wave function 
for both pseudoscalar and vector mesons 
\be\label{rad}
\phi(x_i,{\bf k}_{i\perp})=\frac{4\pi^{3/4}}{\beta^{3/2}}
\sqrt{\frac{\partial k_z}{\partial x}}
{\rm exp}(-{\vec k}^2/2\beta^2),
\ee
where $\beta$ is the variational parameter.
When the longitudinal component $k_z$ is defined by 
$k_z=(x-1/2)M_0 + (m^2_2-m^2_1)/2M_0$, the Jacobian of the variable
transformation $\{x,{\bf k}_\perp\}\to {\vec k}=({\bf k}_\perp, k_z)$
is given by
\be\label{jacob}
\frac{\partial k_z}{\partial x}=\frac{M_0}{4x_1x_2}
\biggl\{ 1- \biggl[\frac{m^2_1-m^2_2}{M^2_0}\biggr]^2\biggr\}.
\ee
Note that the free kinetic part of the Hamiltonian 
$H_0=\sqrt{m^2_q+{\vec k}^2}+\sqrt{m^2_{\bar{q}}+{\vec k}^2}$ 
is equal to the free mass operator $M_0$ in the light-front formalism.

The normalization factor in Eq.~(\ref{rad}) is obtained from the 
following normalization of the total wave function,
\be\label{norm}
\int^1_0dx\int\frac{d^2{\bf k}_\perp}{16\pi^3}
|\Psi^{JJ_z}_{100}(x,{\bf k}_{i\perp})|^2=1.
\ee
Our variational principle to the QCD-motivated effective Hamiltonian
first evaluate the expectation value of the central Hamiltonian $H_0+V_0$,
i.e. $\la\phi|(H_0+V_0)|\phi\ra$ with a
trial function $\phi(x_i,{\bf k}_{i\perp})$ that depends on the
variational parameters $\beta$ and varies $\beta$ until 
$\la\phi|(H_0+V_0)|\phi\ra$ is a minimum.  Once these model
parameters are fixed, then, the mass eigenvalue of each meson is obtained
by $M_{q\bar{q}}=\la\phi|(H_0+V_{q\bar{q}})|\phi\ra$.
On minimizing energies with respect to $\beta$ and searching for a fit
to the observed ground state meson spectra, our central potential
$V_0$ obtained from our optimized  potential parameters 
($a=-0.72$ GeV, $b=0.18$ GeV$^2$, and $\al_s=0.31$)~\cite{CJ1}
for Coulomb plus linear potential were found to be quite comparable with 
the quark potential model suggested by Scora and Isgur~\cite{SI}
where they obtained
$a=-0.81$ GeV, $b=0.18$ GeV$^2$, and $\al_s=0.3\sim 0.6$ for the
Coulomb plus linear confining potential.
More detailed procedure of determining the model parameters of light and
heavy quark sectors can be found in our previous works~\cite{CJ1,CJ2}.
Our model parameters $(m,\beta)$ for the heavy quark sector obtained 
from the linear and HO potential models are summarized in Table~\ref{t1}.
\begin{table}[t]
\caption{The constituent quark mass[GeV] and the Gaussian paramters
$\beta$[GeV] for the linear and HO potentials obtained by the variational
principle. $q=u$ and $d$.}\label{t1}
\begin{tabular}{ccccccccccc} \hline\hline
Model & $m_q$ & $m_s$ & $m_c$ & $m_b$ & $\beta_{qc}$
& $\beta_{sc}$ & $\beta_{cc}$ & $\beta_{qb}$ & $\beta_{sb}$
& $\beta_{bb}$ \\
\hline
Linear & 0.22 & 0.45 & 1.8 & 5.2 &
0.468 & 0.502 & 0.651 & 0.527 & 0.571 & 1.145\\
\hline
HO & 0.25 & 0.48 & 1.8 & 5.2 &
0.422 & 0.469 & 0.700 & 0.496 & 0.574 & 1.803\\
\hline\hline
\end{tabular}
\end{table}

\begin{figure}
\vspace{1cm}
\includegraphics[width=3in,height=3in]{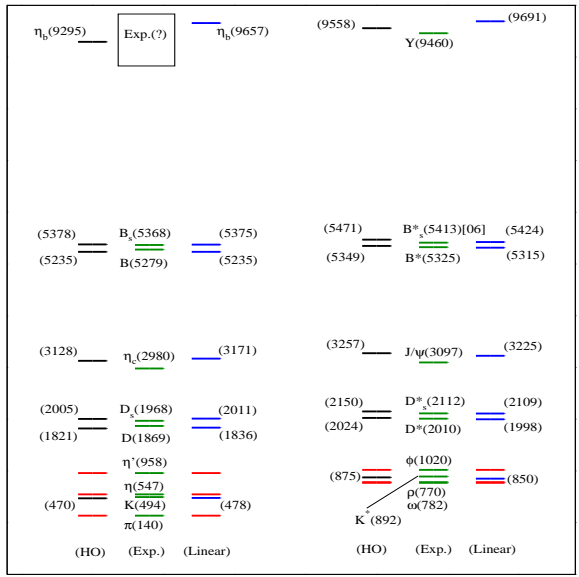}
\caption{(color online).
Fit of the ground state meson masses[MeV] with the parameters given
in Table~\ref{t1}. The $(\rho,\pi)$, $(\eta,\eta')$, and $(\omega,\phi)$ 
masses are our input data. The masses of $(\omega-\phi)$ and $(\eta-\eta')$ 
were used to determine the mixing angles of $\omega-\phi$ 
and $\eta-\eta'$~\cite{CJ1}, respectively.}
\label{fig1}
\end{figure}

Our predictions of the ground state pseudoscalar and vector meson mass
spectra obtained from the linear potential parameters were already shown
in~\cite{CJ2}. In this work, we include the results obtainted from the
HO potential parameters as well and summarize them in Fig.~\ref{fig1}.
Our predictions of the ground state meson mass spectra obtained from both
linear and HO parameters agree with the experimental data~\cite{Data06}
within 6$\%$ error. We should note that our previously predicted mass 
of $B^*_s$, $M_{B^*_s}=5424[5471]$ MeV obtained from the
linear[HO] parameters is in good agreement with the very recent CLEO data,
$M_{B^*}=5414\pm1\pm4$ MeV~\cite{Bon} and 
$M_{B^*}=5411.7\pm1.6\pm0.6$ MeV~\cite{Aqu}.
For the experimentally unmeasured mass of $\eta_b$ meson,
our prediction of mass difference between the two bottomonia
$\Delta m(= M_{\Upsilon}-M_{\eta_b})=34[263]$ MeV obtained from the 
linear[HO] parameters is consistent with current theoretical 
estimates(from perturbative QCD and lattice NRQCD),
$\Delta m=34\sim141$ MeV~\cite{ALEPH}.
As we shall see in our numerical calculations, 
the radiative decay of $\Upsilon\to\eta_b\gamma$ might be useful to 
determine the mass of $\eta_b$ experimentally 
since the decay width $\Gamma(\Upsilon\to\eta_b\gamma)$ is 
very sensitive to the value of $\Delta m$, viz. 
$\Gamma\propto(\Delta m)^3$.

\section{Decay constants and Radiative decay widths}
The decay constants of pseudoscalar and vector mesons are defined by 
\bea\label{fp}
\la 0|\bar{q}\gamma^\mu\gamma_5 q|P\ra&=&if_P P^\mu,
\nonumber\\
\la 0|\bar{q}\gamma^\mu q|V(P,h)\ra&=&f_V M_V\ep^\mu(h),
\eea
where the experimental value of vector meson decay constant $f_V$ 
is extracted from the longitudinal($h=0$) polarization. 
In the above definitions for the decay constants, the experimental values 
of pion and rho meson decay constants are $f_\pi\approx 131$ MeV 
from $\pi\to\mu\nu$ and $f_\rho\approx 220$ MeV from $\rho\to e^+e^-$.

Using the plus component($\mu=+$) of the current, one can easily calculate the
decay constants.
The explicit forms of pseudoscalar and vector meson
decay constants are given by
\bea\label{fp_LFQM}
f_P&=&2\sqrt{6}\int \frac{dx\; d^2{\bf k}_\perp}{16\pi^3}
\frac{{\cal A}}{\sqrt{{\cal A}^2 + {\bf k}^2_\perp}}\phi(x,{\bf k}_\perp),
\nonumber\\
f_V&=&2\sqrt{6}\int \frac{dx\; d^2{\bf k}_\perp}{16\pi^3}
\frac{\phi(x,{\bf k}_\perp)}{\sqrt{{\cal A}^2 + {\bf k}^2_\perp}}
\biggl[ {\cal A} + \frac{2{\bf k}^2_\perp}{{\cal M}_0}\biggr],
\eea
where ${\cal A}=x_2m_1 + x_1m_2$ and ${\cal M}_0=M_0+m_1+m_2$. 

\begin{figure}
\includegraphics[width=3.0in,height=1.5in]{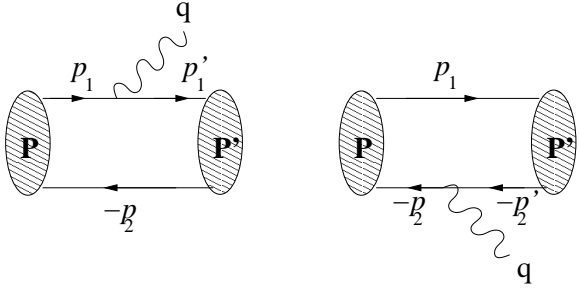}
\caption{Lowes-order graph for $V\to P\gamma^*$ transitions.} 
\label{fig2}
\end{figure}

In our LFQM calculation of $V\to P\gamma$ decay process, we shall first 
analyze the virtual photon($\gamma^*$) decay process so that we calculate
the momentum dependent transition form factor, $F_{VP}(q^2)$. 
The lowest-order Feynman diagram for $V\to P\gamma^*$ process is
shown in Fig.~\ref{fig2} where the decay from vector meson to pseudoscalar
meson and virtual photon state is mediated by a quark loop with flavors
of consituent mass $m_1$ and $m_2$.

The transition form factor $F_{VP}(q^2)$ for the magnetic dipole 
decay of vector meson $V(P)\to P(P')\gamma^*(q)$ is defined as \be\label{ff}
\la P(P')|J^\mu_{\rm em}|V(P,h)\ra
=ie\epsilon^{\mu\nu\rho\sigma}\epsilon_\nu(P,h)
q_{\rho}P_{\sigma}F_{VP}(q^2),
\ee
where the antisymmetric tensor $\ep^{\mu\nu\rho\sigma}$ assures 
electromagnetic gauge invariance, $q=P-P'$ is the four
momentum of the virtual photon, $\ep_\nu(P,h)$ is the polarization 
vector of the initial meson with four momentum $P$ and helicity $h$. 
The kinematically allowed momentum transfer squared $q^2$ ranges 
from 0 to $q^2_{\rm max}=(M_V-M_P)^2$.

The decay form factor $F_{VP}(q^2)$ can be obtained in the 
$q^+=0$ frame with the ``good" component of currents, i.e.
$\mu=+$, without encountering zero-mode contributions~\cite{ZM}.
Thus, we shall perform our LFQM calculation in the $q^+=0$ frame, where
$q^2=q^+q^- - {\bf q}^2_\perp=-{\bf q}^2_\perp<0$, and then analytically
continue the form factor $F_{VP}({\bf q}^2_\perp)$ in the spacelike region
to the timelike $q^2>0$ region by changing ${\bf q}_\perp$ to 
$i{\bf q}_\perp$ in the form factor.

The quark momentum variables for 
$V(q_1\bar{q}_2)\to P(q'_1\bar{q}_2)$ transitions
in the $q^+=0$ frame are given by
\bea\label{mom}
p^+_1&=&x_1 P^+, p^+_2 = x_2 P^+,
\nonumber\\
{\bf p}_{1\perp}&=& x_1{\bf P}_\perp + {\bf k}_\perp,
{\bf p}_{2\perp}= x_2{\bf P}_\perp - {\bf k}_\perp,
\nonumber\\
p'^+_1&=&x_1 P^+, p'^+_2 = x_2 P^+,
\nonumber\\
{\bf p'}_{1\perp}&=& x_1{\bf P'}_\perp + {\bf k'}_\perp,
{\bf p'}_{2\perp}= x_2{\bf P'}_\perp - {\bf k'}_\perp,
\eea
where $x_1=x$ and $x_2=1-x$ and the spectator quark   
requires that $p^+_2=p'^+_2$ and ${\bf p}_{2\perp}={\bf p'}_{2\perp}$.
In the calculations of the decay form factor $F_{VP}(q^2)$, we use
`+'-component of currents and the transverse($h=\pm 1$) polarization.
For the longitudinal($h=0$) polarization, it is hard to extract the form
factor since both sides of Eq.~(\ref{ff}) are vanishing 
for any $q^2$ value.

The hadronic matrix element of the plus current, 
$\la J^+\ra\equiv \la P(P')|J^+_{\rm em}|V(P,h=+)\ra$ in
Eq.~(\ref{ff}) is then obtained by the convolution formula of the initial 
and final state light-front wave functions: 
\bea\label{Jplus}
\la J^+\ra&=&\sum_{j}ee_j\int^1_0\frac{dx}{16\pi^3}
\int d^2{\bf k}_\perp\phi(x,{\bf k'}_\perp)
\phi(x,{\bf k}_\perp)
\nonumber\\
&&\times\sum_{\lam\bar{\lam}}{\cal R}^{00\dagger}_{\lam'\bar{\lam}}
\frac{\bar{u}_{\lam'}(p'_1)}{\sqrt{p'^+_1}}\gamma^+
\frac{u_{\lam}(p_1)}{\sqrt{p^+_1}}
{\cal R}^{11}_{\lam\bar{\lam}},
\eea
where ${\bf k'}_\perp={\bf k}_\perp - x_2{\bf q}_\perp$
and $ee_j$ is the electrical charge for $j$-th quark flavor. 
Comparing with the right-hand-side of Eq.~(\ref{ff}), i.e.
$eP^+F_{VP}(Q^2)q^R/\sqrt{2}$ where $q^R=q_x +iq_y$, 
we could extract the one-loop integral, $I(m_1,m_2,q^2)$,
which is given by 
\bea\label{soft_form}
I(m_1,m_2,q^2) &=&\int^1_0 \frac{dx}{8\pi^3}\int d^2{\bf k}_\perp
\frac{\phi(x, {\bf k'}_\perp)\phi(x,{\bf k}_\perp)}
{x_1\tilde{M_0}\tilde{M'_0}}
\nonumber\\
&&\times
\biggl\{{\cal A} 
+ \frac{2}
{{\cal M}_0}
[{\bf k}^2_\perp
-
\frac{({\bf k}_\perp\cdot{\bf q}_\perp)^2}{{\bf q}^2_\perp}]
\biggr\},
\nonumber\\
\eea
where the primed factors are the functions of final state momenta,
e.g. $\tilde{M_0}'=\tilde{M_0}'(x,{\bf k}'_\perp)$. 

Then, the decay form factor $F_{VP}(q^2)$ is obtained as
\be\label{FS}
F_{VP}(q^2)= e_1I(m_1,m_2,q^2) + e_2 I(m_2,m_1,q^2).
\ee
The coupling constant $g_{VP\gamma}$ for real photon($\gamma$) case
can then be determined in the limit
as $q^2\to 0$, i.e. $g_{VP\gamma}=F_{VP}(q^2=0)$.
The decay width for $V\to P\gamma$ is given by 
\be\label{width}
\Gamma(V\to P\gamma)=\frac{\alpha}{3}g_{VP\gamma}^2 k^3_\gamma,
\ee
where $\alpha$ is the fine-structure constant  and
$k_\gamma=(M^2_V-M^2_P)/2M_V$ is the kinematically allowed energy
of the outgoing photon.

\section{Numerical Results}
In our numerical calculations, we use two sets of model parameters
($m,\beta$) for the linear and HO confining potentials 
given in Table~\ref{t1}
to perform, in a way, a parameter-free-calculation of decay constants 
and decay rates for heavy pseudoscalar and vector mesons.
Although our predictions of ground state heavy meson masses are overall 
in good agreement with the experimental values,
we use the experimental meson masses in the computations of the
radiative decay widths to reduce possible theoretical uncertainties.
But in the case of $\eta_b$, for which experimental data is not yet available,
we use the model mass as $M_{\eta_b}=9353\pm50$ MeV, i.e. we use
slightly broader range
$\Delta m = M_{\Upsilon}-M_{\eta_b}=60-160$ MeV than that
reported in~\cite{ALEPH}.

\begin{table*}[t]
\caption{Charmed meson decay constants(in unit of MeV) obtained
from the linear[HO] parameters.}\label{t2}
\begin{tabular}{ccccccc} \hline\hline
& $f_D$ & $f_{D^*}$ & $f_{D_s}$ & $f_{D^*_s}$ &$f_{\eta_c}$ 
& $f_{J/\psi}$  \\
\hline
Linear[HO] & 211[194] & 254[228] & 248[233] & 290[268] 
& 326[354] & 360[395] \\
\hline
Lattice~\cite{Bec}   & $211\pm14^{+2}_{-12}$
& $245\pm20^{+3}_{-2}$ & $231\pm12^{+8}_{-1}$ 
& $272\pm16^{+3}_{-20}$ & $-$ & $-$ \\
QCD\;\;\;~\cite{Aubin}   & $201\pm3\pm17$
& $-$ & $249\pm3\pm16$ & - & $-$ & $-$ \\
\hline
Sum-rules~\cite{Nar}   & $204\pm20$
& $-$ & $235\pm24$ & - & $-$ & $-$ \\
\hline
BS~\cite{Wang}   & $230\pm25$ & $340\pm23$ & $248\pm27$ 
& $375\pm24$& $292\pm25$ & $459\pm28$ \\ 
\hline
QM~\cite{CG} & $240\pm20$ & $-$ & $290\pm 20$ & $-$ & $-$ & $-$ \\
RQM~\cite{Ebert} & 234 & 310 & 268 & 315 
&$-$ & $-$ \\
\hline
Exp.        & $222.6\pm16.7^{+2.8}_{-3.4}$~\cite{Cleo05} 
& $-$ & $282\pm16\pm7$~\cite{Cleo06} &-& $335\pm 75$~\cite{Cleo_eta} 
& $416\pm6$~\cite{Data06} \\
\hline\hline
\end{tabular}
\end{table*}
In Table~\ref{t2}, we present our predictions for the charmed meson decay
constants($f_D, f_{D^*}, f_{D_s},f_{D^*_s},f_{\eta_c},f_{J/\psi}$) 
together with lattice QCD~\cite{Bec,Aubin}, QCD sum rules~\cite{Nar}, 
relativistic Bethe-Salpeter(BS) model~\cite{Wang}, 
relativized quark model~\cite{CG}, and other 
relativistic quark model(RQM)~\cite{Ebert} predictions as well as 
the available experimental data~\cite{Data06,Cleo05,Cleo06,Cleo_eta}. 
Note that we extract the experimental value
$(f_{J/\psi})_{\rm exp}=(416\pm6)$ MeV from the data 
$\Gamma_{\rm exp}(J/\psi\to e^+e^-)=5.55\pm0.14\pm0.02$ keV~\cite{Data06} 
and the formula~\cite{NR}
\bea\label{qq_de}
\Gamma(V\to e^+e^-)&=&\frac{4\pi}{3}\frac{\al^2}{M_V}f^2_Vc_V,
\eea
where $c_V=4/9$ for $V=J/\psi$. Our predictions for the ratios 
$f_{D_s}/f_D=1.18[1.20]$ and $f_{\eta_c}/f_{J/\psi}=0.91[0.90]$
obtained from the linear[HO] parameters are in good agreement 
with the available experimental data, 
$(f_{D_s}/f_D)_{\rm exp.}=1.27\pm0.12\pm0.03$(preliminary)~\cite{Cleo06}
and $(f_{\eta_c}/f_{J/\psi})_{\rm exp.}=0.81\pm0.19$~\cite{Cleo_eta,Data06},
respectively. Our result for the ratio $f_{D^*_s}/f_{D^*}=1.14[1.18]$
obtained from the linear[HO] parameters is also consistent with the quenched 
lattice result, 1.11(3)~\cite{Bec} and the BS one, 
$1.10\pm0.06$~\cite{Wang}. Overall, our results for the charmed meson 
decay constants are in good agreement with other theoretical model 
calculations as well as the experimental data. 

\begin{table*}[t]
\caption{Bottomed meson decay constants(in unit of MeV) 
obtained from the linear[HO] parameters.}\label{t3}
\begin{tabular}{ccccccc} \hline\hline
 & $f_B$ & $f_{B^*}$ & $f_{B_s}$ & $f_{B^*_s}$ & $f_{\eta_b}$
& $f_{\Upsilon}$  \\
\hline
Linear[HO] & 189[180] & 204[193] & 234[237] & 250[254] 
& 507[897] & 529[983] \\
\hline
Lattice~\cite{Bec}  & $179\pm18^{+34}_{-9}$
& $196\pm24^{+39}_{-2}$ & $204\pm16^{+36}_{-0}$
& $229\pm20^{+41}_{-16}$ & $-$ & $-$ \\
QCD\hspace{0.2cm}~\cite{Gray}  & $216\pm22$ & $-$ & $259\pm32$ 
& $-$ & $-$ & $-$ \\
\hspace{0.9cm}~\cite{Hash}& $189\pm27$ & $-$ & $230\pm30$ & - & $-$ & $-$ \\
\hline
Sum-rules~\cite{Jamin}    & $210\pm19$
& $-$ & $244\pm21$ & - & $-$ & $-$ \\
\hspace{0.9cm}~\cite{Nar}   & $203\pm23$
& $-$ & $236\pm30$ & - & $-$ & $-$ \\
\hline
BS~\cite{Wang}   & $196\pm29$ & $238\pm18$ & $216\pm 32$ & $272\pm20$ 
&$-$ & $498\pm20$ \\
\hline
QM~\cite{CG} & $155\pm 15$ & $-$ & $210\pm 20$ & $-$ & $-$ & $-$ \\
RQM~\cite{Ebert} & 189 & 219 & 218 & 251 
&$-$ & $-$ \\
\hline
Exp.    & $229^{+36+34}_{-31-37}$~\cite{Belle_B} 
& $-$ & $-$ & - & $-$ & $715\pm5$~\cite{Data06} \\
\hline\hline
\end{tabular}
\end{table*}
In Table~\ref{t3}, we show our results for the bottomed meson decay
constants($f_B,f_{B^*},f_{B_s},f_{B^*_s},f_{\eta_b},f_{\Upsilon}$) together 
with lattice QCD~\cite{Bec,Gray,Hash}, QCD sum rules~\cite{Nar,Jamin}, 
BS model~\cite{Wang}, relativized quark model~\cite{CG}, and RQM~\cite{Ebert} 
predictions as well as the available experimental data~\cite{Data06,Belle_B}.
Note that we extract the experimental value 
$(f_{\Upsilon})_{\rm exp}=(715\pm5)$ MeV 
from the data $\Gamma_{\rm exp}(\Upsilon\to e^+e^-)=1.340\pm0.018$ 
keV~\cite{Data06} and Eq.~(\ref{qq_de}) with $c_V=1/9$ for $V=\Upsilon$.
Our results for the ratios $f_{B_s}/f_{B}=1.24[1.32]$ 
and $f_{B^*_s}/f_{B^*}=1.23[1.32]$ obtained from the 
linear[HO] parameters are quite comparable with the recent lattice results, 
$1.20(3)(1)$~\cite{Gray} and $1.22(^{+5}_{-6})$~\cite{Hash} 
for $f_{B^*_s}/f_{B^*}$ and
$1.17(4)^{+1}_{-3}$~\cite{Bec} for $f_{B^*_s}/f_{B^*}$.
For the $\Upsilon$ meson decay constant, our
prediction $f_{\Upsilon}= 529[893]$ MeV  obtained from the linear[HO]
parameters slightly deviates from the extracted experimental value 
$(f_{\Upsilon})_{\rm exp}=(715\pm5)$ MeV. Other model calculations 
for $f_{\Upsilon}$ such as $498\pm20$ MeV from the BS model~\cite{Wang} and
836 MeV from effective Lagrangian satisfying heavy-quark spin 
symmetry(HQSS)~\cite{HQSS} also show some deviations 
from the experimental value. Our result for the ratio 
$f_{\eta_b}/f_{\Upsilon}= 0.96[0.91]$ obtained from the linear[HO] 
parameters is to be compared with the $f_{\eta_b}/f_{\Upsilon}\sim 1$ 
in HQSS limit~\cite{HQSS}. For these heavy bottomed meson decay constants,
we observe an overall agreement between our results and other theoretical
ones.

\begin{table}[t]
\caption{Coupling constants $g_{VP\gamma}$[GeV$^{-1}$]
for radiative $V\to P\gamma$ decays obtained from the linear[HO] 
parameters.}\label{t4}
\begin{tabular}{cccccc} \hline\hline
Coupling & This work & ~\cite{Jaus96} & ~\cite{Barik}
&  ~\cite{GI} & Exp.~\cite{Data06}\\
\hline
$g_{J/\psi\eta_c\gamma}$
& 0.681[0.673] & - & - & 0.69 & $0.57\pm0.11$ \\
\hline
$g_{D^{*+}D^+\gamma}$ & -0.384[-0.398] & -0.30 & -0.37 & -0.35
&-$(0.50\pm0.12)$\\
\hline
$g_{D^{*0}D^0\gamma}$ & 1.783[1.826] & 1.85 & 1.94 & 1.78 & -\\
\hline
$|\frac{g_{D^{*0}D^0\gamma}}{g_{D^{*+}D^+\gamma}}|$
& 4.64[4.59] & 6.17 & 5.24 & 5.08 & -\\
\hline
$g_{D^{*+}_sD^+_s\gamma}$ & -0.167[-0.161] & - & -0.17 & -0.13 &- \\
\hline
$g_{B^{*+}B^+\gamma}$ & 1.311[1.313] & 1.40 & 1.50 & 1.37 & - \\
\hline
$g_{B^{*0}B^0\gamma}$ & -0.749[-0.750] &-0.80 & -0.85 & -0.78 & - \\
\hline
$|\frac{g_{B^{*0}B^0\gamma}}{g_{B^{*+}B^+\gamma}}|$
& 0.57[0.57] & 0.57 & 0.57 & 0.57 & - \\
\hline
$g_{B^{*0}_sB^0_s\gamma}$ & -0.553[-0.536]& - & -0.62 & -0.55 & -\\
\hline
$g_{\Upsilon\eta_b\gamma}$ & -0.124[-0.119]& - & - & -0.13 & -\\
\hline\hline
\end{tabular}
\end{table}
In Table~\ref{t4}, we present our results of the coupling constants 
$g_{VP\gamma}$(in unit of GeV$^{-1}$) for radiative $V\to P\gamma$ decays 
together with other QM calculations~\cite{Jaus96,Barik,GI} 
as well as the available experimental data. 
The experimental values for $(g_{J/\psi\eta_c\gamma})_{\rm exp}=0.57\pm0.11$
for $J/\psi\to\eta_c\gamma$ and 
$(g_{D^{*+}D^+\gamma})_{\rm exp}=-(0.50\pm0.12)$
for $D^{*+}\to D^+\gamma$ processes are extracted from the
branching ratios  
${\rm Br}(J/\psi\to\eta_c\gamma)_{\rm exp}=(1.3\pm 0.4)\%$ and
${\rm Br}(D^{*+}\to D^+\gamma)_{\rm exp}=(1.6\pm 0.4)\%$ together with 
the full widths of $\Gamma_{\rm tot}(J/\psi)=93.4\pm 2.1$ keV
and $\Gamma_{\rm tot}(D^{*+})=96\pm 22$ keV~\cite{Data06}.
The opposite sign of coupling constants for $D^{*+}$ and
$D^{*+}_s$ decays compared to the charmonium $J/\psi$ decay indicates
that the charmed quark contribution is largely destructive in the 
radiative decays of $D^{*+}$ and $D^{*+}_s$ mesons.
Similarly, we see that the bottomed quark contribution is largely 
destructive in the radiative decay of $B^{*+}$ meson.
Our predictions for $g_{J/\psi\eta_c\gamma}=0.681[0.673]$
and $g_{D^{*+}D^+\gamma}=-0.384[-0.398]$ obtained from the linear[HO] 
parameters fall within the experimental error bars. 
Our result for the coupling constant ratio 
$|\frac{g_{D^{*0}D^0\gamma}}{g_{D^{*+}D^+\gamma}}|=4.64[4.59]$ 
obtained from the linear[HO] parameters
is quite comparable with other theoretical model predictions such
as those $6.32\pm2.97$~\cite{Dosc} and $3.05\pm 0.63$~\cite{Aliev} from
the QCD sum rules, $5.54\pm3.00$~\cite{HQET} from the heavy quark effective 
theory(HQET), and $4.49\pm0.96$~\cite{Thews} from the
broken-SU(4) symmetry by M1 transition. Incidentally, our result for 
the coupling constant ratio
$|\frac{g_{B^{*0}B^0\gamma}}{g_{B^{*+}B^+\gamma}}|=0.57[0.57]$ obtained
from the linear[HO] parameters is the same as that from the other QM 
predictions~\cite{Jaus96,Barik,GI}. This result is also comparable with
$0.64\pm0.51$~\cite{Dosc} and $0.49\pm 0.38$~\cite{Aliev}
from the QCD sum rules, and $0.59\pm0.48$~\cite{HQET} from the HQET.

\begin{figure*}[t]
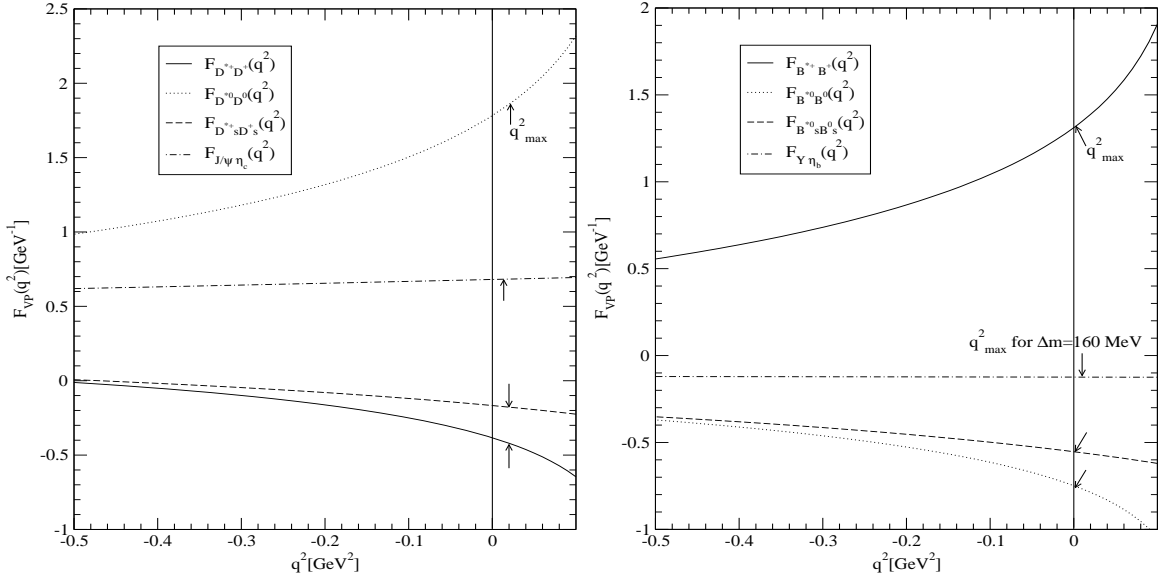

\vspace{1cm}
\includegraphics[width=3.0in,height=3.0in]{fig3a.eps}
\includegraphics[width=3.0in,height=3.0in]{fig3b.eps}
\caption{Transition form factors $F_{VP}(q^2)$ for charmed(left panel)
and bottomed(right panel) mesons radiative decays obtained from the
linear parameters.}
\label{fig3}
\end{figure*}
We show in Fig.~\ref{fig3} our results of decay form factors 
$F_{VP}(q^2)$ obtained from the linear parameters. 
Since the results from the HO parameters are not much different 
from those of linear ones, we omit them for simplicity. 
The left(right) panel shows the results of charmed(bottomed)
vector meson radiative $V\to P\gamma^*$ decays.
The solid, dotted, dashed, and dot-dashed lines in the left(right)
panel represent the form factors for
$D^{*+}\to D^+\gamma^*(B^{*+}\to B^+\gamma^*)$, 
$D^{*0}\to D^0\gamma^*(B^{*0}\to B^0\gamma^*)$,
$D^{*+}_s\to D^+_s\gamma^*(B^{*0}_s\to B^0_s\gamma^*)$,
and
$J/\psi\to\eta_c\gamma^*(\Upsilon\to\eta_b\gamma^*)$ decays,
respectively. The arrows in the figure represent the zero recoil points
of the final state pseudoscalar meson, i.e.  $q^2=q^2_{\rm max}=(M_V-M_P)^2$.
We have performed the analytical continuation of the decay form
factors $F_{VP}(q^2)$ from the spacelike region($q^2<0$) to the physical 
timelike region $0\leq q^2\leq q^2_{\rm max}$.
The coupling constant $g_{VP\gamma}$ at $q^2=0$ corresponds
to a final state pseudoscalar meson recoiling with maximum 
three-momentum $|\vec{P}_P|=(M^2_V-M^2_P)/2M_V$ in the rest frame 
of vector meson.
Due to the small kinematic region $0\leq q^2\leq q^2_{\rm max}$ for
the bottomed and bottomonium meson decays, the recoil effects of the
final state mesons are quite negligible, i.e.  
$F_{VP}(q^2_{\rm max})/g_{VP\gamma}\approx 1$. Likewise, we find that
$F_{J/\psi\eta_c}(q^2_{\rm max})/g_{J/\psi\eta_c\gamma}
\approx F_{D^*_s D_s}(q^2_{\rm max})/g_{D^*_s D_s\gamma}\approx 1$ for
$J/\psi\to\eta_c\gamma^*$ and $D^{*+}_s\to D^+_s\gamma^*$ decays.
On the other hand, we obtain 
$F_{D^{*+} D^+}(q^2_{\rm max})/g_{D^{*+} D^+\gamma}
=0.420/0.384\approx 1.1$ and
 $F_{D^{*0} D^0}(q^2_{\rm max})/g_{D^{*0} D^0\gamma}
=1.859/1.783\approx 1.04$
for $D^{*+}\to D^{*+}\gamma^*$ and $D^{*0}\to D^{*0}\gamma^*$decays, 
respectively. The recoil effect, i.e. the difference between the
zero($q^2_{\rm max}$) and the maximum($q^2=0$) points, may not be
negligible especially for the $D^{*+}\to D^+\gamma^*$ decay.
Fig.~\ref{fig3} also shows the restoration of SU(3) flavor symmetry, 
$F_{D^{*+}_s D^{*+}_s}(q^2)/F_{D^{*+}D^{*+}}(q^2)\to 1$ between charmed
and charmed-strange mesons
and $F_{B^{*0}_s B^{*0}_s}(q^2)/F_{B^{*0}B^{*0}}(q^2)\to 1$
between bottomed and bottomed-strange mesons
in the intermediate and deep spacelike($q^2<0$) region, where the
light quark current contribution becomes negligible.

\begin{table*}[t]
\caption{Decay widths and branching ratios 
for radiative $V\to P\gamma$ decays obtained from our
linear[HO] model parameters. 
We used $M_{\eta_b}=9353\pm50$ MeV for $\Upsilon\to\eta_b\gamma$
decay.}\label{t5}
\begin{tabular}{cccc} \hline\hline
Decay mode & $\Gamma$[keV] & 
$\rm Br$ &  Br$_{\rm exp}$~\cite{Data06} \\
\hline
$J/\psi\to\eta_c\gamma$ 
& $1.69\pm0.05[1.65\pm0.05]$ 
&$(1.80\pm 0.10)[1.76\pm 0.10]\%$ & $(1.3\pm 0.4)\%$ \\
\hline
$D^{*+}\to D^+\gamma$ & $0.90\pm0.02[0.96\pm0.02]$ 
&$(0.93\pm0.31)[1.00\pm 0.34]\%$ & $(1.6\pm 0.4)\%$ \\
\hline
$D^{*0}\to D^0\gamma$ & $20.0\pm0.3[21.0\pm0.3]$
&- & $(38.1\pm 2.9)\%$ \\
\hline
$D^{*+}_s\to D^+_s\gamma$ & $0.18\pm0.01[0.17\pm0.01]$
&- & $(94.2\pm 0.7)\%$ \\
\hline
$B^{*+}\to B^+\gamma$ & $0.40\pm0.03[0.40\pm0.03]$
&- & $-$ \\
\hline
$B^{*0}\to B^0\gamma$ & $0.13\pm0.01[0.13\pm0.01]$
&- & $-$ \\
\hline
$B^{*0}_s\to B^0_s\gamma$ 
& $0.068\pm0.017[0.064\pm0.016]$
&- & $-$ \\
\hline
$\Upsilon\to\eta_b\gamma$ 
& $0.045^{+0.097}_{-0.038}[0.042^{+0.088}_{-0.036}]$ 
& $(8.4^{+18.6}_{-7.2})[7.7^{+17.0}_{-6.6}]\times10^{-4}$ & $-$ \\
\hline\hline
\end{tabular}
\end{table*}
For a more direct comparison with the available experimental data, we 
finally calculate the partial decay widths from Eq.~(\ref{width}). In 
Table~\ref{t5}, we present our results for the decay  widths and 
branching ratios together with the available experimental data. 
The errors in our results for the decay widths and branching ratios
come from the uncertainties of the experimental mass 
values and experimental mass values plus the full widths, respectively. 
Our results of the branching ratios
${\rm Br}(J/\psi\to\eta_c\gamma)=1.80\pm 0.10[1.76\pm 0.10]\%$
and
${\rm Br}(D^*\to D^+\gamma)= 0.93\pm0.31[1.00\pm0.34]\%$
obtained from the linear[HO] parameters are in agreement with the
experimental data~\cite{Data06}, 
${\rm Br}(J/\psi\to\eta_c\gamma)_{\rm exp}=(1.3\pm 0.4)\%$
and
${\rm Br}(D^*\to D^+\gamma)_{\rm exp}=(1.6\pm 0.4)\%$ within the error bars. 
For the neutral charmed meson decay, our prediction
$\Gamma(D^{*0}\to D^0\gamma)=20.0\pm0.3[21.0\pm 0.3]$ keV
obtained from the linear[HO] parameters is to be compared with other 
theoretical model results such as 21.69 keV from the RQM~\cite{Jaus96},
14.40 keV from the QCD sum rules~\cite{Aliev} and
$27.0\pm 1.8$ keV from broken-SU(4) symmetry by M1 transition~\cite{Thews}.
For the charmed-strange meson decay, our prediction
$\Gamma(D^{*+}_s\to D^+_s\gamma)=0.18\pm0.01[0.17\pm 0.01]$ keV
obtained from the linear[HO] parameters is comparable with other
theoretical model results such as 0.19 keV from the RQM~\cite{EFG1} and
0.3 keV~\cite{Cheng} and $(0.24\pm0.24)$ keV~\cite{HQET} from the HQET.
Since the $D^{*0}$ lifetime has not been measured
yet, we also try to estimate the full width for $D^{*0}$ meson
using the relation
\bea\label{DR}
\frac{{\rm Br}(D^{*+}\to D^+\gamma)}{{\rm Br}(D^{*0}\to D^0\gamma)}
=\frac{\Gamma(D^{*+}\to D^+\gamma)}{\Gamma(D^{*0}\to D^0\gamma)}
\frac{\Gamma_{\rm tot}(D^{*0})}{\Gamma_{\rm tot}(D^{*+})},
\eea
where we use our predicted decay width $\Gamma(D^{*0}\to D^0\gamma)$ to
extract the full width for $D^{*0}$.
Similarly, we can estimate the full width for $D^{*+}_s$ meson
using the same method as in the case of $D^{*0}$ meson. 
Our averaged values of the full widths for $D^{*0}$ and
$D^{*+}_s$ mesons obtained from the two parameter sets are
\bea\label{width_D}
\Gamma_{\rm tot}(D^{*0})&=&(55\pm 6)\; {\rm keV},
\nonumber\\
\Gamma_{\rm tot}(D^{*+}_s)&=&(0.19\pm 0.01)\;{\rm keV},
\eea
respectively, while experimentally only upper limits were reported
as $\Gamma(D^{*0})_{\rm exp}<2.1$ MeV and
$\Gamma(D^{*+}_s)_{\rm exp}<1.9$ MeV. 
Some other theoretical model predictions of the full widths for
$D^{*0}$ and $D^{*+}_s$ mesons were also reported as 
$\Gamma_{\rm tot}(D^{*0})=65.09$ keV from the RQM~\cite{Jaus96}
and $\Gamma_{\rm tot}(D^{*0})=(36.7\pm 9.7)$ keV 
and $\Gamma_{\rm tot}(D^{*+}_s)=(0.24\pm0.24)$ keV
from the HQET~\cite{HQET}.

For $B^{*}$ and $B^{*}_s$ radiative decays, our results for the
decay widths 
$\Gamma(B^{*+}\to B^+\gamma)=0.40\pm0.03[0.40\pm0.03]$,
$\Gamma(B^{*0}\to B^0\gamma)=0.13\pm0.01[0.13\pm0.01]$, 
and $\Gamma(B^{*0}_s\to B^0_s\gamma)=0.068\pm0.017[0.064\pm0.016]$
obtained from the linear[HO] parameters
are quite comparable with other theoretical model
predictions such as $\Gamma(B^{*+}\to B^+\gamma)=0.429$ keV
and $\Gamma(B^{*0}\to B^0\gamma)=0.142$ keV
from the RQM~\cite{Jaus96}, $\Gamma(B^{*+}\to B^+\gamma)=(0.22\pm0.09)$ keV
and $\Gamma(B^{*0}\to B^0\gamma)=(0.075\pm0.027)$ keV
from the HQET~\cite{HQET}, and $\Gamma(B^{*+}\to B^+\gamma)=0.14$ keV
and $\Gamma(B^{*0}\to B^0\gamma)=0.09$ keV
from the chiral perturbation theory~\cite{Amu}.
Finally, for the $\Upsilon\to\eta_b\gamma$ process, our predictions for the
decay width and branching ratio obtained
from the linear[HO] parameters are 
$\Gamma(\Upsilon\to\eta_b\gamma)
=45^{+97}_{-38}[42^{+88}_{-36}]\;{\rm eV}$ and
${\rm Br}(\Upsilon\to\eta_b\gamma)
=(8.4^{+18.6}_{-7.2})[7.7^{+17.0}_{-6.6}]\times 10^{-4}$,
where the upper, central, and lower values correspond to $\Delta m=60$ MeV,
110 MeV, and 160 MeV, respectively. For this bottomonium radiative decay, 
the decay width $\Gamma(\Upsilon\to\eta_b\gamma)$ is found to be very 
sensitive to $\Delta m$ because it is proportional to $(\Delta m)^3$. 
Our result is to be compared with other model predictions such as
$\Gamma(\Upsilon\to\eta_b\gamma)=(3.6\pm2.9)$ eV~\cite{Nora} from
the nonrelativistic effective field theory model,
$(33.2\pm0.1)$ eV~\cite{Hwang} and 5.8 eV~\cite{EFG2} from the RQM.

\section{Summary and Discussion}
In this work, we investigated the weak decay constants and the
magnetic dipole $V\to P\gamma$ decays
of heavy-flavored mesons such as $(D,D^*,D_s,D^{*}_s,\eta_c, J/\psi)$ 
and $(B,B^*,B_s,B^*_s,\eta_b,\Upsilon)$ using the LFQM
constrained by the variational principle for the QCD-motivated effective
Hamiltonian. The momentum dependent form factors $F_{VP}(q^2)$ for
$V\to P\gamma^*$ decays are obtained
in the $q^+=0$ frame and then analytically continued to the timelike region
by changing ${\bf q}_\perp$ to $i{\bf q}_\perp$ in the form factors. 
The coupling constants $g_{VP\gamma}$, which are needed for the calculations 
of the decay widths for $V\to P\gamma$, can then be determined in the limit 
as $q^2\to 0$, i.e.  $g_{VP\gamma}=F_{VP}(q^2=0)$.
Our model parameters obtained 
from the variational principle uniquely determine the above nonperturbative 
quantities. This approach can establish the extent of applicability of our 
LFQM to wider ranging hadronic phenomena.

Our predictions of mass spectra and decay constants for heavy pseudoscalar 
and vector mesons are overall in good agreement with the available 
experimental data as well as other theoretical model calculations. 
Our numerical results
of the decay widths for $J/\psi\to\eta_c\gamma$ and $D^{*+}\to D^+\gamma$
fall within the experimental error bars. We also estimates the unmeasured
full widths for $D^{*0}$ and $D^{*+}_s$ as 
$\Gamma_{\rm tot}(D^{*0})=(55\pm 6)$ keV and
$\Gamma_{\rm tot}(D^{*+}_s)=(0.19\pm 0.01)$ keV, respectively.  
Our predictions for the branching ratios for the bottomed and 
bottomed-strange mesons are quite comparible with other theoretical model 
predictions.
For the radiative decay of the bottomonium, we find that the decay widths 
$\Gamma(\Upsilon\to\eta_b\gamma)$ is very sensitive to the value of 
$\Delta m=M_\Upsilon-M_{\eta_b}$. This sensitivity for the
bottomonium radiative decay may help to determine the mass 
of $\eta_b$ experimentally. In going beyond the static result to 
see the momentum dependence of the form factor for $V\to P\gamma^*$,
we find that most results in the heavy flavored sector
stand almost unaffected from the recoil effects. 
However, the form factor $F_{D^{*+}D^+}(q^2)$ seems to give
a non-negligible recoil effect about $10\%$ between zero and maximum
recoil points, i.e. 
$F_{D^{*+}D^+}(q^2_{\rm max})/g_{D^{*+}D^+\gamma}\approx 1.1$.

Since the form fator $F_{VP}(q^2)$ of vector meson radiative decay 
$V\to P\gamma^*$ presented in this work is precisely analogous to the 
vector current form factor $g(q^2)$ in weak
decay of ground state pseudoscalar meson to ground state vector 
meson, the ability of our model to describe such decay is 
therefore relevant to the reliability of the model for
the weak decay. Consideration on such exclusive weak decays
in our LFQM is underway. Although our previous LFQM~\cite{CJ1,CJ2} and
this analyses did not include the heavy mesons comprising both 
$c$ and $b$ quarks such as $B_c$ and $B^*_c$, the extension of
our LFQM to these mesons will be explored in our future communication. 

\acknowledgments
This work was supported by a grant from Korea
Research Foundation under the contract KRF-2005-070-C00039.

\end{document}